# Experimental TDPAC and Theoretical DFT Study of Structural, Electronic, and Hyperfine Properties in ($^{111}$In→) $^{111}$Cd-Doped SnO$_2$ Semiconductor: *Ab Initio* Modeling of the Electron-Capture-Decay After-Effects Phenomenon.

Germán N. Darriba[1,*], Emiliano L. Muñoz[2], Artur W. Carbonari[3], and Mario Rentería[1,†]

[1] *Departamento de Física and Instituto de Física La Plata (IFLP, CONICET-UNLP), Facultad de Ciencias Exactas, Universidad Nacional de La Plata, CC 67, 1900 La Plata, Argentina.*
[2] *Facultad de Ingeniería, Universidad Nacional de La Plata, 1900 La Plata, Argentina.*
[3] *Instituto de Pesquisas Energéticas e Nucleares-IPEN/CNEN, Universidade de São Paulo, São Paulo, SP, Brazil.*

In this paper we investigate the effect of Cd doping at ultra-low concentrations in SnO$_2$ both experimentally, by measuring the temperature dependence of the electric quadrupole hyperfine interactions with time-differential perturbed angular correlation (TDPAC) spectroscopy using $^{111}$Cd as probe nuclei, and theoretically, by performing first-principles calculations based on the density functional theory. TDPAC spectra were successfully analyzed with a time-dependent *on-off* model for the perturbation factor. These results show combined dynamic plus static interactions whose electric-field gradients were associated in this model to different stable electronic configurations close to the Cd atoms. The dynamic regime is then originated in fast fluctuations between these different electronic configurations. First-principles calculations results show that the Cd impurity introduces a double acceptor level in the top of the valence band of the doped semiconductor and produces isotropic outward relaxations of the nearest oxygen neighbors. The variation of the calculated electric-field gradient tensor as a function of the charge state of the Cd impurity level shows an interesting behavior that explains the experimental results, giving strong support from first-principles to the *electron-capture after-effects* proposed scenario. The electron-capture decay of the parent $^{111}$In to $^{111}$Cd as well as the double acceptor character of the $^{111}$Cd impurity and the electric nature of the host are shown to contribute to the existence of these type of time-dependent hyperfine interactions.

I. INTRODUCTION

The wide-bandgap SnO$_2$ semiconductor is a very attractive material for a wide variety of technological applications, due to its electrical and optical properties as well as its thermal and chemical stability. It has been used in the construction of solar cells,[1] flat panel collectors, touch panels[2] and sensors for gas detection.[3,4] It is well known that the inclusion of certain impurities in SnO$_2$ can improve these properties.[5-8] More recently, Cd has revealed as an ideal dopant for some of these properties. For example, in the case of gases sensors, the Cd doping (i.e., the Sn$^{+4}$ substitution by Cd$^{+2}$) induces the formation of oxygen vacancies, enhancing the sensor performance with respect to undoped n-type SnO$_2$ (in which natural oxygen vacancies may be present in the real samples) at the same operating temperature.[9] On the other hand, the inclusion of Cd impurities in SnO$_2$ nanoparticles has revealed a dielectric constant loss behavior at high frequencies[10] suggesting its application in many electronic devices. Concerning optoelectronic applications, Mariappan *et al.*[11] developed Cd-doped SnO$_2$ thin films enhancing their electrical conductivity and optical transmittance when 5% of Cd doping is used.

In the last years, in condensed-matter physics, structural, electronic, and magnetic properties in pure and doped systems have been carefully studied at the atomic scale confronting results from hyperfine experimental techniques with reliable all-electron (AE) *ab initio* predictions in the framework of the Density Functional Theory (DFT).[12-20] In particular, the subnanoscopic environment of impurities or native atoms in solids can be studied employing the Time-Differential Perturbed γ−γ Angular-Correlation (TDPAC) spectroscopy. This technique provides a very precise characterization of the electric-field-gradient (EFG) tensor at diluted (ppm) radioactive probe-atoms. The EFG is a symmetric tensor of second-order with components defined by $V_{ij}(\vec{r}) = \frac{\partial^2 V(\vec{r})}{\partial x_i \partial x_j}$. Here, $V(\vec{r})$ is the electric potential generated by the charge density nearby a probe nucleus. Particularly, the EFG tensor is diagonal in the principal axis system, and may be characterized by only its largest component $V_{33}$ and the asymmetry parameter defined as $\eta = (V_{11} - V_{22})/V_{33}$, with $0 \leq \eta \leq 1$ using the $|V_{11}| \leq |V_{22}| \leq |V_{33}|$ standard convention. In summary, from the determination of the EFG tensor it is possible to obtain very valuable information about the impurity-host system under study, such as structural distortions, localization of impurities and defects as well as their charge state, etc., and this information could be obtained by confrontation of the experimental results with accurate theoretical predictions of the EFG.

As it is well known, the ($^{111}$In→) $^{111}$Cd and ($^{181}$Hf→)$^{181}$Ta nuclides are the most used probe-atoms in TDPAC spectroscopy, both as native atom or as an impurity in a given system. In the case of $^{181}$Hf(→$^{181}$Ta)-doped $SnO_2$ a complete study combining TDPAC experiments and *ab initio* calculations has been performed in detail.[15] This double approach enabled us to determine the structural relaxations introduced in the host, the correct charge state of the impurity and the metallic behavior of a degenerate semiconductor (in agreement with resistivity experimental results obtained in samples with the same Ta dilution), and we could understand the origin of the EFG invariance with temperature observed in the TDPAC measurements.

On the other hand, for the $^{111}$In(→$^{111}$Cd)-doped $SnO_2$ system, the existing TDPAC experiments reported in the literature [21-24] are not conclusive in what respect to the interpretation of the hyperfine parameters reported, and their temperature dependence remains controversial. Besides the samples preparation and their activation, one of the difficulties in the analysis of these TDPAC spectra resides in the fact that the transmutation of $^{111}$In to $^{111}$Cd is produced by nuclear electron capture (EC) ($^{111}$In(EC) →$^{111}$Cd), giving rise to the electronic relaxation process usually called in the literature (electron-capture-decay) "after-effects" (ECAE).[25-27]

The first TDPAC results of $^{111}$Cd-doped $SnO_2$ were reported by Wolf *et al.*[21] In their experiments, the $^{111}$In(→$^{111}$Cd) probe´s parent was diffused into $SnO_2$ powder samples and measured in the temperature (T) range 4K-1260 K. They found a unique hyperfine interaction (characterized by $\omega_Q$=18.2(1) Mrad/s and $\eta$=0.1(1), temperature independent), splitted into a *static* and a *dynamic* part. They found a temperature dependence of the fraction ($f$) of probes that senses the static part of this interaction, varying from 25% at T=4 K to 70% at T > 1000 K. They attributed this dependence to the ECAE and to the fact that the perturbation depends of the local electronic density, and proposed the generation of unfilled inner electron shells to explain this temperature dependence. Giving support to these ideas, they observed a faster increase of $f$ as T increases when the $SnO_2$ samples were doped with "donor" $H_2$ atoms. The presence of a unique hyperfine interaction was consistent with the single nonequivalent Sn site of the rutile structure. Bibiloni *et al.*[22] performed TDPAC experiments diffusing $^{111}$In(→$^{111}$Cd) atoms into pure Sn foils (and then oxidizing the activated samples to obtain $^{111}$In-doped $SnO_2$) and in high purity $SnO_2$ powders. For both kinds of samples two static hyperfine interactions were observed (characterized at room temperature (RT) by $\omega_{Q1}$=31.1(9) Mrad/s, $\eta_1$=0.45(4) and $\omega_{Q2}$=14.3(6) Mrad/s, $\eta_2$=0.65), with their populations varying as a function of T in the 17 K-1173 K range. These interactions were assigned to $^{111}$Cd probe-atoms located at substitutional Sn sites with singly and doubly ionized oxygen vacancies as near neighbors, respectively. Subsequently, similar hyperfine interactions governed by the same temperature dependence were observed by Moreno *et al.*[23] diffusing $^{111}$In($\rightarrow^{111}$Cd) probes into $SnO_2$ thin films (employing two different backing substrates). Bibiloni *et al.* and Moreno *et al.* did not find in their investigations any evidence of the existence of a time-dependent interaction arising from the ECAE. Finally, Rentería *et al.*[24] performed TDPAC experiments at RT in air on $^{111}$In($\rightarrow^{111}$Cd)-implanted Sn-O films after each step of thermal annealings up to 1023 K. Initially, they found two coexistent phases, disorder SnO and $SnO_2$, which were transformed by an appropriate annealing treatment into the $SnO_2$ crystalline phase. Four hyperfine interactions were needed to reproduce the TDPAC spectra, two corresponding to $^{111}$Cd in SnO and the other two to $^{111}$Cd in $SnO_2$. The hyperfine parameters of $^{111}$Cd in the crystalline $SnO_2$ phase were $\omega_Q$=18.4(1) Mrad/s and $\eta$=0.18(2), in agreement with the hyperfine parameters found by Wolf *et al.*, but without evidence of a time-dependent interaction. It should be mentioned that this EFG is also in agreement with one of the hyperfine interactions observed by Bibiloni *et al.*,[22] which was only observed at T=1073 K.

Taking into account all of these previous and dissimilar experimental results and their interpretations and in order to solve this controversy, we performed new and careful TDPAC experiments in high purity $SnO_2$ polycrystalline samples. With the aim to maximize the percentage of $^{111}$Cd probes located in defect-free Sn sites, we diffused $^{111}$In($\rightarrow^{111}$Cd) into $SnO_2$ powders under a non-oxidant $N_2$ atmosphere. The TDPAC measurements were carried out as a function of temperature, in order to study the ECAE phenomenon. In addition, we performed a complete *ab initio*/DFT study that gives a reliable description of structural and electronic effects introduced by the impurity in this semiconductor oxide, without the use of external parameters and self-consistently. We performed this detailed study as a function of the charge state of the impurity, which enables us to present an *ab initio* scenario to interpret the ECAE phenomenon.

In what follows, we present the TDPAC technique, the data reduction, the sample preparation, and the time-dependent perturbation factor used to analyze the experimental results, which are shown in Section II. In Section III we present the first-principles procedure and discuss the corresponding theoretical predictions for the EFG and for the electronic and structural properties of Cd-doped $SnO_2$. In Section IV we discuss the experimental and the *ab initio* results, presenting an *ab initio* model for the ECAE and, finally, in Section V we present our conclusions.

## II. EXPERIMENTAL

### A. TDPAC Spectroscopy, Data Reduction, and Time-Dependent Perturbation Factor

TDPAC spectroscopy is based on the conservation of the angular momentum and basically it consists in the determination of the perturbation (generated by extranuclear fields) on the correlation between the emission directions of two successive $\gamma$ radiations ($\gamma_1$ and $\gamma_2$) during a nuclear decay. This perturbation occurs during the lifetime of the intermediate state of the $\gamma_1$-$\gamma_2$ cascade. For a complete description of this spectroscopy, see Refs. 28-30.

In a TDPAC experiment we measure the number of coincidences $C(\Theta,t)$ between $\gamma_1$ and $\gamma_2$ photons detected at two relative angular positions of the scintillators, $\Theta=90°$ and $\Theta=180°$, being $t$ the elapsed time between both emissions. Hence, the experimental spin-rotation curve or $R(t)$ TDPAC spectrum can be constructed as:

$$R(t) = 2\frac{C(180°)-C(90°)}{C(180°)+2C(90°)} \cong A_{22}^{exp} G_{22}^{exp}(t) \qquad (1)$$

where $A_{22}^{exp}$ is the experimental anisotropy of the $\gamma_1$-$\gamma_2$ cascade and $G_{22}^{exp}(t)$ is the theoretical perturbation factor $G_{22}(t)$ folded with the time-resolution curve of the spectrometer.

The TDPAC experiments presented here were performed using the 171-245 keV $\gamma$-$\gamma$ cascade of $^{111}$Cd, produced after the EC nuclear decay of the parent $^{111}$In nuclide.

In what follows we will show the construction and justification of the perturbation factor used to describe the time-dependent hyperfine interactions observed in the TDPAC experiments reported in the present work. This analytic perturbation factor was proposed by Bäverstam *et al.*[31] (BO) and is a modification of that proposed by Abragam and Pound (AP) used to describe the randomly fluctuating dynamic (i.e., time-dependent) hyperfine interaction sensed by a probe-nucleus in a liquid environment.[32] In the BO model, a modification was introduced in order to account for the *on-off* character of the dynamic hyperfine interaction, character that is related to the probe-atom probability to decay to a certain stable electronic ground state (turning *off* the dynamic interaction), during the time-window of the TDPAC measurement, leading to a final static interaction.

The type of dynamic interaction we want to describe here is originated in the radioactive decay of the parent of the TDPAC probe if this decay leaves the probe atom in a highly ionized state far from the equilibrium. This occurs, e.g., in the K-electron-capture decay of $^{111}$In to $^{111}$Cd, probe used in this work. This process produces an electronic hole in the K shell of the Cd atom, which is filled up with an electron of the outer shells too fast to have any influence on the TDPAC spectra. After this, and due to subsequent Auger processes, the probe ion loses a large number of electrons, leaving the corresponding number of electronic holes, which will diffuse towards the outer atomic shells in less than $10^{-12}$ s. This first part of the atomic recovery is independent of the environment of the probe atom and again does not influence the TDPAC spectra. However, the following phase during which the outer holes are filled up depends on the probe surroundings. In a solid, the highly ionized charge state of the Cd probe-atom becomes almost compensated by electrons from their neighboring ions in a very short time ($\approx 10^{-12}$ s). In the electronic band structure description of the solid, these holes diffuse along the valence band very fast. This happens before the emission of the first $\gamma$-ray of the $\gamma_1$-$\gamma_2$ cascade through which the $^{111}$Cd nucleus decays to its (nuclear) ground state (the emission of $\gamma_1$ starts the TDPAC measurement and that of $\gamma_2$ ends it).

Only ions with very few *extra* electronic holes with respect to those nominally introduced by a neutral Cd atom substituting a host cation (in our case, i.e. Cd-doped SnO$_2$, two holes are nominally produced) may live for times long enough (longer than $10^{-12}$ s) to reach the sensitive time-window of the measurement. The necessary condition for this phenomenon to take place seems to be that the probe-atom should be embedded in an insulating or semiconducting environment.[21,25,33-39] During the time-window of the measurement, transitions among all the resulting potential probe´s charge states (including a final stable charge state) can occur and the probe nucleus will feel fluctuating EFGs due to the transitions among different electronic configurations of the probe atom and its nearest neighbors.

Differently to other proposals that lead to numerical simulations of the *R(t)* spectrum (see, e.g., Ref. 40), one of the advantages of the BO model consists that it leads to an analytical expression for the perturbation factor that can be used for the spectra fitting. Let's describe now in more detail the construction of the perturbation factor in the BO model[40] used to analyze the TDPAC results in this work.

The most general expression for the perturbation factor $G_{kk}(t)$ may be written as:

$$G_{kk}(t) = \sum_i c_i e^{a_i t}, \qquad (2)$$

where $c_i$ and $a_i$ may be complex numbers.

In the case of *static* (time-independent) electric-quadrupole interactions, i. e. when the electronic environment of the probe-nucleus does not change during the time interval defined by the emission of $\gamma_1$ and $\gamma_2$, the perturbation factor, $G_{22}^s(t)$, for nuclear-electric-quadrupole interactions, polycrystalline samples and spin $I=5/2^+$ of the intermediate nuclear level of the $\gamma_1$-$\gamma_2$ cascade (as in the case of $^{111}$Cd) is:

$$G_{22}^s(t) = S_{20} + \sum_{n=1}^{3} S_{2n}(\eta) \cos(\omega_n(\eta) t) e^{-\delta \omega_n t}, \qquad (3)$$

where the coefficients $S_{2n}$ and the interaction frequencies $\omega_n$ are known functions of the asymmetry parameter $\eta$, being $\omega_n$ proportional to the nuclear quadrupole interaction frequency $\omega_Q = eQV_{33}/40\hbar$.[41] In Eq. 3, we used $k=2$ since $A_{44}$ is very small for this $\gamma$-$\gamma$ cascade.

In the case that the following conditions are satisfied (as may be the case of randomly fast fluctuating electronic environments sensed by the TDPAC probe):

(i) the correlation time $\tau_c$, which is a typical time for the changes of the nuclear environments that will originate a time-dependent interaction, is small enough to let $\omega_Q \ll 1/\tau_c$ ;

(ii) the time of observation t $\gg \tau_c$;

(iii) the fluctuating interaction is so small that the first-order perturbation theory is valid;

then a *pure dynamic* (i. e., time-dependent) perturbation factor is obtained as:

$$G_{22}^{dyn}(t) = e^{-t/\tau_r} = e^{-\lambda_r t}, \qquad (4)$$

and $\tau_r(=1/\lambda_r)$ is called the relaxation time constant. In our case atomic recombination and electronic de-excitation time constants are included in $\tau_r$. These processes have very short correlation times $\tau_c$. Thus conditions (i) and (ii) seem to be fulfilled. In our case, as also occurred in Ref. 31, it is sufficient to fit the $R(t)$ spectra with only one exponential as in Eq. 4, showing that the fluctuating interaction is small. On other hand, if the fluctuating interaction is stronger than what is assumed in (iii), $G_{22}^{dyn}(t)$ will be the sum of two exponentials, according to the Dillenburg and Maris theory.[42]

The shape of the TDPAC spectra in our experiments (yet observed in many TDPAC experiments in semiconducting and insulating oxides using [111]In, see Refs. 25-27,43, and 44) suggests that the nuclei of the excited [111]Cd atoms are feeling a static electric-field gradient (namely that corresponding to a certain [111]Cd final electronic stable state) besides a dynamic hyperfine interaction (represented by the exponential factor of Eq. 4). Since, in addition, the static interaction is much weaker than the dynamic one (i.e. the exponential decay is strong, this can be seen in the fast damping of the spectra), $G_{22}(t)$ can be well described, for the *combined* static and dynamic interactions, as the product:

$$G_{22}(t) = G_{22}^{s}(t)e^{-\lambda_r t}. \qquad (5)$$

This equation represents a static perturbation factor with a time-dependent anisotropy which decreases to zero for very large times. This is not exactly the shape of our R(t) spectra, which remains undampened after a certain time *t*. The fact that the excited atoms after a certain time will reach their electronic ground states (these states may be different at each temperature), i.e. the turning *off* of the dynamic interaction, had not yet been accounted for in Eq. 5. To do this, two simplifying assumptions were made:

(i) That the probability for an atom reaching its ground electronic state at time $t'$ (measured from the emission time of $\gamma_1$) is:

$$P_g(t') = \lambda_g N e^{-\lambda_g t'}, \qquad (6)$$

where $N$ is the number of probes that will contribute to $G_{22}(t)$ at time $t$ (those probes that emits $\gamma_2$ at time $t$) and $\lambda_g$ is the probability per unit time for an atom to reach the electronic ground state.

(ii) That the mean (dynamic) interaction strength averaged over all excited atoms is constant, i.e., $\lambda_r$ is still a constant.

After the $N$ atoms reach their electronic ground state (at a time $t'$, different for each atom), the hyperfine interaction they produce becomes static. Thus $G_{22}(t)$ may now be seen as an average of the perturbation factors affecting to the *N* probe-nuclei. The contribution to the perturbation factor, at time *t*, corresponding to a probe that arrives at the electronic ground state at time *t′*, is (remember that $\gamma_2$ is emitted at time *t*):

$$G_{22}(t,t') = \begin{cases} G_{22}^S(t)e^{-\lambda_r t'} & \text{if } t > t' \\ G_{22}^S(t)e^{-\lambda_r t} & \text{if } t < t' \end{cases}. \qquad (7)$$

This situation is graphically described in Fig. 1 that shows the contribution to the total time-dependent $G_{22}(t)$ perturbation factor from probes that reach stable electronic state at a time *t′*. In this figure, $G_{22}(t)$ was multiplied by the experimental anisotropy $A_{22}$ to obtain the *R*(*t*) spectrum.

One way to obtain the total time-dependent perturbation factor $G_{22}(t)$ at a time *t* is to compute the average of Eq. 7, taken on the *N* atoms that emit $\gamma_2$ at time *t*, including those atoms that reach their electronic ground state at time *t′* before *t* and those that are still excited at time t (i.e., *t′* is larger than *t* in Eq. 7) :

$$G_{22}(t) = N^{-1} \int_0^\infty G_{22}(t,t') P_g(t') dt' \qquad (8)$$

$$G_{22}(t) = N^{-1} \left( \int_0^t G_{22}^S(t)e^{-\lambda_r t'} \lambda_g N e^{-\lambda_g t'} dt' + \int_t^\infty G_{22}^S(t)e^{-\lambda_r t} \lambda_g N e^{-\lambda_g t'} dt' \right), \qquad (9)$$

then

$$G_{22}(t) = \left[ \frac{\lambda_r}{\lambda_r+\lambda_g} e^{-(\lambda_r+\lambda_g)t} + \frac{\lambda_g}{\lambda_r+\lambda_g} \right] G_{22}^S(t). \qquad (10)$$

This model cannot be distinguished from the physically different case where there are *two fractions* of probes, $f_d$ undergoing a fluctuating field (without the *off* process) characterized by the perturbation factor:

$$G_{22}(t) = G_{22}^S(t)e^{-\lambda t} \qquad (11)$$

and $f_s$ subjected only to the static field. Indeed, Eq. 10 is equivalent to:

$$G_{22}(t) = [f_d e^{-\lambda t} + f_s] G_{22}^S(t), \qquad (12)$$

if $f_d=\lambda_r/\lambda$ , $f_s=\lambda_g/\lambda$ , with $\lambda=\lambda_r+\lambda_g$. In Eq. 12, $f_d$, $f_s$, and $\lambda$ are, in general, not correlated parameters. This is not the case, clearly, in the scenario represented by Eq. 10, which will be used to fit the *R*(*t*) spectra.

In case the probe-atoms in a sample are localized at different environments, e.g. at nonequivalent cation sites of a crystal structure, Eq. 1 and Eq. 10 need to be modified accordingly using the usual model for multiple-site electric-quadrupole- interactions for the static perturbation factor:

$$R(t) \cong A_{22}^{exp} G_{22}^{exp}(t) = A_{22}^{exp} \sum_i f_i G_{22_i}^{exp}(t) =$$

$$A_{22}^{exp} \sum_i f_i \left[ \frac{\lambda_{r_i}}{\lambda_{r_i}+\lambda_{g_i}} e^{-(\lambda_{r_i}+\lambda_{g_i})t} + \frac{\lambda_{g_i}}{\lambda_{r_i}+\lambda_{g_i}} \right] G_{22_i}^{exp}(t), \qquad (13)$$

being $f_i$ the fraction of nuclei that senses each time-dependent hyperfine interaction.

### B. Sample preparation and TDPAC measurements

The sample was prepared using high purity rutile $SnO_2$ powder (Sigma-Aldrich, purity better than 99.99 %), pressed in a circular pellet of 5 mm diameter at $3 \times 10^8$ Pa, and annealed in air for 1h at 423 K, plus 6 h at 773 K and 10 h at 1073 K in order to sinter and improve its crystallinity. The $^{111}$In($\rightarrow^{111}$Cd) isotope was introduced afterwards by thermal diffusion, dropping approximately 10 μCi of $^{111}$InCl$_3$ dissolved previously in 0.05 normal HCl solution in water. This pellet was placed in a sealed quartz tube under $N_2$ atmosphere at low pressure ($2 \times 10^4$ Pa) and annealed at 1073 K for 12 h.

The TDPAC spectrometer used was based on four $BaF_2$ scintillating detectors in a 90° coplanar arrangement and an electronic coincidence set-up with slow-fast logic. The $R(t)$ spectrum was constructed for each measuring temperature, according to Eq. 1 from the twelve simultaneously recorded coincidence spectra, eight at 90° and four at 180°.

TDPAC measurements were carried out in the temperature range from 10 K to 1123 K in a reversible way, in 50 K or 25 K steps. For measurements above RT, the sample was sealed in an evacuated quartz tube and performed in a small furnace with graphite electrodes positioned between detectors. For low-temperature measurements in the range from 10 K to 295 K, the sample was attached to the cold finger of a closed-cycle helium cryogenic device with temperature controlled to better than 0.1 K.

### C. Experimental Results

In Fig. 2 we show selected representative $R(t)$ spectra and their corresponding Fourier transformation measured at T= 10 K, 150 K, 200 K, 270 K, 450 K, 650 K, 923K, and 1123 K. As it is apparent, there is an increasing damping as measurement temperature decreases, starting from the very well defined and undamped spectrum at T=1123 K. This behaviour is reversible with the measurement temperature.

Already at 1123 K two hyperfine interactions (corresponding to two different triplets of interaction frequencies) are resolved in the Fourier spectrum, although one of them (HFI1) accounts for almost 95 % of the spectrum. Below 923 K a fast damping in the first ns of the $R(t)$ spectra is apparent but conserving this anisotropy for larger times. As mentioned before this effect increases as temperature decreases. Hence, as explained in Sec. II A, the (red) solid lines in the $R(t)$ spectra of Fig. 2 are the best least-squares fit of Eq. 13 to the experimental data. The (red) solid lines in the Fourier spectra are the Fourier transformation of the $R(t)$ fits.

Two hyperfine interactions (HFI1 and HFI2) were necessary to account for the $R(t)$ spectra along all the temperature range of measurement. The contribution of HFI1 and HFI2 are shown as shaded areas (in light-blue and blue, respectively) in the Fourier spectra (Fig. 2). The fitted more relevant hyperfine parameters are shown in Fig. 3 as a function of temperature. In this figure, the fitting errors on $\omega_Q$ and $\eta$ are smaller than the data points. In the case of the fractions $f$, the absolute errors are about ±5% and the relative errors in $\lambda_r$ and $\lambda_g$ are about 10 % (not included in this figure for clarity). As mentioned in Sec. II A, the $\lambda_r$ relaxation constant represents the damping strength in a pure time-dependent interaction (see Eq. 4) related with the transitions among different probe´s charge states, being $\lambda_g$ the inverse of the mean lifetime of the "electronic holes", $\tau_g$, which governs the *"on-off time"* of the dynamic interaction. In addition, Fig. 4 shows the relative weights of the *dynamic* and *static* terms of Eq. 12 for HFI1 and HFI2.

The combined inspection of the evolution of the hyperfine parameters (Fig. 3), the dynamic and static fraction extracted from Eq. 12 (Fig. 4) and the contributions of HFI1 and HFI2 to the Fourier spectra (Fig. 2) enable us to describe the scenario sensed by the $^{111}$Cd probe-atoms as a function of temperature. Above 923 K the spectra present a pure static behaviour, with more than 80% of probes sensing HFI1, characterized at T=1123 K by $\omega_{Q1}$=17.9(2) Mrad/s, $\eta_1$=0.13(1), in agreement with the values of the single hyperfine interaction observed by Wolf et al.[21] While $f_1$ varies with temperature as seen in Fig. 3(a), $\omega_{Q1}$ and $\eta_1$ are almost constant as well as its EFG distribution ($\delta$ < 2 %). This constancy was also observed by Wolf et al.[21] and is in agreement with the characterization performed at RT by Rentería et al.[24] As Fig. 4 (upper) shows, $f_s$ of HFI1 (solid light-blue squares) equals 100% from 1123 K to 650 K and then decreases as T decreases with a step-like behavior. This is in agreement with the temperature behavior of the "static fraction" of the single hyperfine interaction reported by Wolf *et al.*[21] On the other hand, $f_s$ of HFI2, characterized along all the temperature range of measurement by $\omega_{Q2}$ =21-24 Mrad/s and $\eta_2$=0.2-0.6 (see Fig. 2 (d)-(e)), is 100% only from 1123 K to 923 K (see lower part of Fig. 4). Between 923 K and 650 K only HFI2 presents a dynamic behaviour, see Figs 3(b), 3(c) and Fig. 4 (lower). As these figures show, none of both interactions has a pure static behaviour for T lower than 650 K.

At this point, the results obtained applying the *"on-off"* B-O model for the $R(t)$ of Eq. 13 let us propose the following scenario for the $^{111}$Cd probe-atoms. At high T (T ≥ 923 K), all the probes feel a pure static EFG, with more than 90% of them sensing an EFG reflecting the same final stable electronic state (EFG1) reached after the *"turn off"* of the dynamic process, process that is observed for HFI1 only below 650 K. The rest of the probes (10%) sense a temperature dependent EFG corresponding to different final stable charge states of the $^{111}$Cd impurity [see $\omega_{Q2}$ in Fig. 3(e)], reached after the *"turn off"* of the dynamic process for each temperature (process only observed below 923 K in the case of HFI2). At medium and low T, the probe senses initially a *combined* static and time-dependent interaction (see Eq. 5) related with randomly fluctuating EFGs originated in the transitions between different electronic configurations of the probe-atom (different charge states of $^{111}$Cd including the corresponding final stable charge state related with EFG1 or EFG2, characterizing HFI1 or HFI2, respectively) until the *"turn off"* of this process. After this, the probe senses a pure static interaction (characterized by EFG1 or EFG2) related with the final stable electronic charge state the probe has reaches at each temperature. It is useful to point out again that EFG1 is the same for all temperatures while EFG2 is temperature dependent [see Fig. 3(e)].

On the other hand, the fractions $f_1$ and $f_2$ of probes that feel the final EFG1 and EFG2, respectively, change with temperature [see Fig. 3(a)]. As T decreases, $f_2$ increases and HFI2 begin to sense the dynamic interaction (below 923 K), while $f_1$ decreases, being HFI1 still a static interaction (only above 650 K). View to the fact that $f_1$ increases with temperature and reaches more than 90 % above 923 K, temperature region with a larger electron availability due to thermal effects, and being EFG1 the same for all temperatures, it is reasonable to suppose that EFG1 is related with a probe´s charge state with more electrons than those associated with EFG2.

In general, as T decreases, the electron availability and/or mobility could be decreasing in the system, as suggested by the increase of $\lambda_r$ and the constancy of $\lambda_g$ (i.e., with large values for $\tau_g$) at low T [see Figs. 3(b) and 3(c)]. In effect, the increase of $\lambda_r$ reflects an increase of the strength of the random fluctuation among different electronic configurations due to a decreasing electron availability, as T decreases. On the other hand, large $\tau_g$ values could be understood in terms of low electron availability and mobility. Following this idea, the fraction evolution of HF1 and HF2 from high temperature to RT could be explained by a decrease of the available electrons of the host, i.e. the relative lacking of electrons as T decreases produces a decrease of the number of probes that senses EFG1 and an increase of those atoms that sense the interaction related with EFG2 (which need less electrons than EFG1, as we supposed above). At T lower than RT, $f_1$ unexpectedly increases again. But at the same time both $\lambda_{r1}$ and $\lambda_{r2}$ strongly increases (especially $\lambda_{r2}$) while $\tau_{g1}$ and $\tau_{g2}$ remain relatively constant at very large values [see Figs. 3 (b) and (c)]. Both effects, more electronic holes with large lifetimes giving rise to both time-dependent interactions, produce the charge required to increase the fraction of probes that feel EFG1 as final state.

Finally, based in the constancy of $\omega_{Q1}$ and $\eta_1$ and the low value of $\delta_1$ along all the temperature range of measurement, their agreement with the values reported by Wolf *et al.* for the single interaction observed in their experiments, and the static character of the EFG1 along a wide high temperature range, we tentatively assigned HFI1 to $^{111}$Cd probes localized at cation sites of the SnO$_2$ crystalline structure. While HFI1 should be correlated with the same final charge state of the impurity, the oscillating behaviour and large EFG2 distribution suggest that HFI2 should be correlated with slightly different final electronic configurations of $^{111}$Cd also localized at Sn sites. In our scenario, as mentioned before, the probe´s electronic configuration giving rise to EFG1 should have more electrons than the different configurations that produce the temperature dependant values of EFG2. This assumption will be discussed in detail with the help of the *ab initio* study of the impurity-host system, described in what follows.

## III. *AB INITIO* CALCULATIONS

### A. Calculations Details

In order to enlighten the experimental TDPAC results described in the precedent section we performed electronic structure DFT-based *ab initio* calculations in Cd-doped rutile $SnO_2$. To simulate an isolated impurity, as is the case of $^{111}Cd$ atoms doped in our TDPAC samples with ppm impurity dilution, we used a supercell (SC) size such as each impurity do not interact with the nearest ones, and that the structural relaxations of its neighbours do not affect the relaxations of other neighbours of the closest impurities. This SC (2x2x3 SC) is formed by 12unit cells of rutile $SnO_2$. This unit cell is tetragonal with $a=b=4.7374(1)$ Å and $c=3.1864(1)$ Å,[45,46] and it contains two Sn atoms at positions 2*a*: (0;0;0) and (½,½,½), and four O atoms at positions 4*f*: ±($u,u$,0) and ±(½+$u$,½-$u$,½), with $u=0.3056(1)$ [45] or $0.3064(4)$.[46] In this structure, Sn has an octahedral oxygen coordination with four oxygen atoms (O2) forming the octahedron basal plane and two O1 atoms in the vertex of the octahedron [see Fig. 5(a)]. The resulting 2x2x3 SC is tetragonal with lattice parameters $a'=2a=b'=2b=9.47$ Å and $c'=3c=9.56$ Å, obtaining $c'/a'=1.01$. In this SC, one of the 24 Sn atoms is replaced by a Cd one [Fig. 5(b)], obtaining $Sn_{0.958}Cd_{0.042}O_2$, with the shortest distance between Cd impurities of about 9.5 Å. We studied the convergence of the structural relaxations and the hyperfine parameters in other SCs formed by 2x2x2 and 2x2x4 unit cells, checking that the 2x2x3 SC was sufficient to simulate the isolated impurity, as usually occurs in oxides with impurity-impurity shortest distance of around 10 Å.[13,15,18]

In order to describe real samples, and particularly in our case where an electronic recombination process around the impurity occurs, it is essential to be able to predict with accuracy the charge state of the impurity, and therefore to describe correctly the structural, electronic, and hyperfine properties of the impurity-host system. In this sense, we performed calculations for different charge states of the impurity taking into account the nominal double acceptor character of $Cd^{2+}$ when it replaces a $Sn^{4+}$ in the $SnO_2$ host. For this, we add electrons to the SC (once a Sn atom is replaced by a Cd atom) in 0.1 $e^-$ steps and up to 2 $e^-$, and we will call $SnO_2$:$Cd^{x-}$ the SC when x electrons are added to the $SnO_2$:Cd system.

The *ab initio* calculations have been performed employing the Full-Potential Augmented Plane Waves plus local orbitals (FP-APW+lo) method[47] embodied in the WIEN2k code.[48] The cut-off parameter of the plane wave bases in the interstitial region was $R_{MT}K_{max}$=7, where $K_{max}$ is the maximum modulus of the lattice vectors in the reciprocal space, and $R_{MT}$ is the smallest radius of the non-overlapping muffin-tin spheres centered in the atoms. In our calculations we use $R_{MT}$(Cd)= 1.06 Å, $R_{MT}$(Sn)= 0.95 Å and $R_{MT}$(O)= 0.86 Å. In the reciprocal space, the integration was performed using the tetrahedron method,[49] taking 50 *k*-points in the first Brillouin zone. Exchange and correlation effects were treated using the local-density approximation (LDA),[50] and the generalized gradient approximation (GGA). For GGA we used the Perdew–Burke–Ernzerhof (PBE-GGA) parametrization[51] and the GGA parametrization proposed by Wu and Cohen (WC-GGA).[52] In order to obtain the equilibrium structures of the SC, all atoms were displaced following a Newton dampened scheme,[53] and repeating it until the forces on the ions were below 0.01 eV/Å. Finally, the EFG tensor was calculated from the second derivative of the obtained full electric potential.[54, 55]

### B. Calculation Results

#### *1. Structural relaxation*

The substitution of a lattice native Sn atom by a Cd impurity induces considerable large forces acting on Cd and its neighbouring atoms, as well as, in a lesser extent, on the rest of the atoms of the SC. In the Cd-doped 2x2x3 SC, we checked if the final equilibrium positions obtained after the relaxation process depend on the initial structure (lattice and internal parameters) or not. We started the relaxation process using both the experimental and the *ab initio* optimized structural parameters of pure $SnO_2$ (extensively studied in Ref. 15), obtaining almost the same final structure. This can be explained by the enormous relaxation of Cd´s nearest oxygen neighbours positions with respect to the subtle atomic position refinement obtained in the structural optimization process in pure $SnO_2$. What is relevant for our study is that the atomic equilibrium final positions of the neighboring atoms of Cd result the same, hence the electronic configuration and the hyperfine properties at the Cd site do not depend on the initial positions used in the calculations.

In Table I we compare the Cd-O1 and Cd-O2 bond-lengths, obtained after the equilibrium positions are achieved in $SnO_2$:$Cd^0$, $SnO_2$:$Cd^{1-}$ and $SnO_2$:$Cd^{2-}$, to the Sn-O1 and Sn-O2 bond-lengths in pure $SnO_2$. As we see, the differences between the predicted Cd-O1 and Cd-O2 bond-lengths using LDA, WC-GGA, and GGA approximations are less than 0.3 %. For all charge states, the ONN (O1 and O2) atoms relax outwards enlarging the Cd-ONN distance. These dilatations are isotropic and depend strongly on the charge state of the impurity, increasing as electrons are added to the system. The magnitude and sense (outward or inward) of the bond-length relaxations introduced by the Cd impurity when it replaces a Sn atom are related with the difference between their ionic radii, that are 0.69 Å and 0.95 Å for six-fold coordinated $Sn^{4+}$ and $Cd^{2+}$, respectively.[56] In effect, in Ref. 15 it was shown that exist a positive proportionality between the ONN bond-length relaxations and the ionic radii of Ta and Cd impurities relative to the radius of the native cation replaced by the impurities in rutile oxides. Moreover, the oxygen relaxations follow a general trend already observed in Cd-doped binary oxides with different crystal structures, in which the Cd impurity tries to reconstruct the Cd-ONN bond-length of 2.35 Å that it has in its most stable oxide CdO.[57] The largest relaxations were found for the $SnO_2$:$Cd^{2-}$ SC due to the additional coulomb repulsion originated when two electrons are added. Nevertheless, this relaxation does not reach the "ideal" value 2.35 Å that Cd has in CdO since the presence of the nearest Sn atoms of each ONN limits the Cd-ONN relaxation.

*2. Electronic Structure*

In addition to the structural relaxations, the substitution of a Sn atom by a Cd impurity produces strong changes in the electronic structure of the semiconductor. Figs. 6(a) and 6(b) show the density of electronic states (DOS) for pure $SnO_2$ and $SnO_2$:$Cd^0$, respectively, after the equilibrium atomic positions are reached. Comparing these DOS one can see that the Cd impurity introduces an acceptor impurity level at the top of the valence band (VB), basically formed by Cd-*d* and ONN-*p* states. These states contribute also to the bottom and the top of the VB, respectively. This situation is better shown in the atom-projected partial DOS (PDOS) of Fig. 6(c). On the right side of Figs. 6(b) and 6(c), magnifications of the DOS in the energy range of the acceptor impurity level are also shown. Integration of the empty states within the impurity level until the Fermi level (i. e., E=0 eV) for the $SnO_2$:$Cd^0$ system [see right side of Fig. 6(b)] gives a value of 2 $e^-$, confirming the double acceptor character of the $Cd^{2+}$ impurity replacing a $Sn^{4+}$ in $SnO_2$. In Fig. 7, the PDOS of Cd-*d*, O1-*p* and O2-*p* for $SnO_2$:$Cd^0$, $SnO_2$:$Cd^{1-}$, and $SnO_2$:$Cd^{2-}$ are shown for the impurity level energy region. As we can see, when additional charge is added to the *neutral* system ($SnO_2$:$Cd^0$) up to 2 electrons, these charge goes to the Cd impurity and the ONN atoms, filling only states with symmetries $d_z^2$, $d_{x^2-y^2}$ and $d_{xz}$ of Cd, $p_z$ of O1, and $p_x$ and $p_y$ of the O2 atoms (see center and right columns in Fig. 7), in agreement with the coordination geometry of the cationic site in the rutile structure. We have demonstrated that the added electrons are localized in the Cd-O1 and Cd-O2 bonds. In effect, the right side of Fig. 7 shows the local coordinate systems used in the calculations in which the Cd-$d$ and O-$p$ orbitals are referred. According to this, O1-$p_z$ orbitals points along the Cd-O1 bond, and the hybridization of O2-$p_x$ and O2-$p_y$ orbitals along the Cd-O2 bonds, being these bonds shown in Fig. 8. This figure shows the electron density, in the (1-10) and (110) lattice planes containing Cd and O1 and Cd and O2 atoms, respectively. Figs. 8(a) and 8(b) correspond to the electron density for the first electron added to the $SnO_2$:$Cd^0$ system, Figs. 8(c) and 8(d) correspond to the second electron added to the $SnO_2$:$Cd^{1-}$ system, and Figs. 8(e) and 8(f) correspond to the two electrons added to the $SnO_2$:$Cd^0$ system. To obtain the electron densities of Fig. 8, the calculations were performed for selected-occupied states in the impurity level energy region of the $SnO_2$:$Cd^{1-}$ system for Figs. 8(a) and 8(b), and $SnO_2$:$Cd^{2-}$ system for Figs. 8(c) and 8(d). In the case of Figs. 8(e) and 8(f) we use the occupied states corresponding to the last two electrons below the Fermi energy in $SnO_2$:$Cd^{2-}$ system. Comparing Figs. 8(a) and 8(b), the first electron added to the SC has a preference to be localized at Cd and the O1 atoms along the Cd-O1 bonds, whereas the second electron added prefers to be localized at Cd and the O2 atoms along the Cd-O2 bonds. This is due in part because once added the first electron, the coulombian electronic repulsion with the second electron along the Cd-O1 axis is stronger. With respect to the $d$-charge which results localized at the Cd impurity when one or two electrons are added to the system, the second bumps of the radial wave function of the Cd-orbitals in Fig. 8 helps to identify the symmetries dominant in each plane. In Fig. 8(a) Cd-$d_z^2$ and $d_{x^2-y^2}$ orbitals are present while the hybridization of Cd-$d_z^2$ and $d_{x^2-y^2}$ can be detected in Fig. 8(b) (depleted charge along the x axis), as well as an additional contribution of Cd-$d_{xz}$ (see also the local coordinate axes systems at Cd shown in Fig. 7). As mentioned before, the second electron has mainly Cd-$d_{xz}$ character [see Fig. 8(d)] while the hybridization of Cd-$d_z^2$ and $d_{x^2-y^2}$ is more pronounced, as can be deduced from the depleted charge region along the $x$ axis.

### 3. EFG

With the aim to explain the origin of the EFG tensors that characterize the two hyperfine interactions observed in the TDPAC experiments and the potential EFG dependence with the charge state of the impurity that could give account of the dynamic character of these interactions, we carried out EFG calculations at Cd sites in $SnO_2$ as a function of the impurity charge state. The EFG tensors calculated for the three charge states of the SCs $SnO_2$:$Cd^0$, $SnO_2$:$Cd^{1-}$, and $SnO_2$:$Cd^{2-}$, taking into account the full structural relaxation produced in each system, are quoted in Table II. Little differences exist between $V_{33}$ and $\eta$ predicted by LDA, WC-GGA, and PBE-GGA approximations. These differences are due to the little difference in the final equilibrium positions predicted by each approximation (see Table I), due to the strong $r^{-3}$ EFG dependence from the charge sources. Due to this $r^{-3}$ dependence, the EFG at the Cd site is basically originated in the non-spherical electronic charge density very close to the Cd nucleus (in the first bumps of the Cd´s radial wave functions).[55] It is important to notice here that, as occurred in many other doped binary oxide systems,[15,18,58,59] these approximations predict the same EFG tensor (within the precision of the calculations) when the same structural positions are employed. The $V_{33}$ direction is parallel to the [1-10] crystal axis (see Figs. 5 and 8) for all these charge states and is the same that it has in pure $SnO_2$ and Ta-doped $SnO_2$ at Sn and Ta sites, respectively.[15] While $V_{33}$ slightly increases and decreases when 1 and 2 electrons are added to the SC, respectively, the $\eta$ parameter that describes the symmetry of the charge density around the probe nucleus changes drastically once the second electron is added. In order to understand this drastic change and the unexpected $V_{33}$ slight variation, we performed a systematic calculation of the EFG varying, between 0 and 2 electrons, the negative charge added to the *neutral* SC, in steps of 0.1 $e^-$. Fig. 9 shows the result of these calculations for the LDA approximation. In it, the asymmetry parameter and the $p$ and $d$ contributions to the total $V_{ii}$ principal components of the EFG are plotted as a function of the added charge. This study enables the detection of an increase in $V_{33}$ and $\eta$ as electrons are added to the SC and an abrupt decrease of their values for the charge states close to the complete ionization of the impurity level, i.e. when this level is almost filled. While the $p$ contribution to $V_{ii}$ is almost constant for different charge states (and dominant for $V_{33}$ as already observed in other Cd-doped binary oxides),[13,60] the $d$ contribution changes strongly, leading to the shape of the total $V_{ii}$ dependence with the impurity charge state. The strong decrease of $\eta$ towards an axially symmetric situation ($\eta$ =0.14) is explained by the strong variation of $V_{11}$ and $V_{22}$, governed again by the variation of the $d$ contribution. In effect, $V_{11}$ and $V_{22}$ are almost equal when 2 electrons are added to the SC, leading to $\eta$ close to zero.

The constancy of the *p* contribution to $V_{ii}$ as a function of the impurity charge state (from 0 to 2 electrons) can now be understood by the fact that the *p*-states of Cd are almost only present in the VB, whereas the *d*-states are present both in the VB and in the impurity level. Considering, roughly speaking, that the VB states do not change when the impurity level is being filled (*rigid-band* model), the behaviour of the EFG should be described by the change of the relative occupation of Cd-*d* orbitals with $d_{x^2-y^2}$, $d_{xz}$, and $d_{z^2}$ symmetries. A closer inspection of the electronic densities in the neighbourhood of the Cd nucleus plotted at selected planes containing Cd and its ONN (see Fig. 8), which corresponds directly to the above mentioned Cd-*d* orbital symmetries within the acceptor impurity level, could explain the behaviour of the EFG as a function of the impurity charge state (Fig. 9). For this description, we will take into account that a negative point charge along a certain axis produces a negative EFG component in this direction, while if this charge is localized in a plane normal to this direction it produces a positive EFG component with half of the precedent value. In general, the sign and strength of the EFG component along a certain axis originated by a negative charge q situated at an angle Φ from this axis is given by:

$$V_{ii}(\vec{r}) = \frac{q(3\cos^2 \Phi - 1)}{4\pi\varepsilon_0 r^3}. \quad (14)$$

Hence, negative point charges localized between 0 and 54.7° from a certain principal *i* axis produces a negative $V_{ii}$ value.

In effect, as described in detail in the previous section, the addition of one electron to the $SnO_2$:$Cd^0$ SC produces a much larger localization of electronic charge along the Cd-O1 bonds than the charge added along the Cd-O2 bonds. Hence, as mentioned above, the addition of this negative charge along the $V_{22}$ direction produces a positive increase in $V_{33}$, since the $V_{33}$ direction is normal to $V_{22}$. This effect continues increasing $V_{33}$ up to its maximum value and then abruptly decreases when the acceptor impurity level is completely filled. In effect, the second added electron goes mainly to the Cd-O2 bonds [see Figs. 8(c) and 8(d)] producing an electronic configuration in which the relative amount of charge along $V_{22}$ is smaller with respect to the charge distributed in the *x-z* plane (see the local coordinate axes system at Cd in Fig. 7, on the right). This charge, which corresponds mainly to Cd-$d_{xz}$ orbitals, is close to 45° from the $V_{33}$ axis [see Fig. 8(d)], producing a negative contribution to $V_{33}$ (according to Eq. 14), hence decreasing the positive $V_{33}$ value (see Fig. 9, 1st panel).

Following the same ideas, the addition of the first electron along mainly the $V_{22}$ direction increases the negative $V_{22}$ value (from -2 to -5 $\times 10^{21}$ V/m$^2$, see Fig. 9) and the addition of the second electron decreases this value (from -5 to -1), since the additional charge in the *x-z* plane normal to $V_{22}$ originates a positive EFG contribution to $V_{22}$. In the same way, the addition of the second electron that goes mainly to the *x-z* plane contributes with a negative EFG along the *z* axis, decreasing the $V_{11}$ positive value (see Fig. 9, 3$^{th}$ panel).

## IV. COMPARISON BETWEEN EXPERIMENTAL TDPAC RESULTS AND THE *AB INITIO* CALCULATIONS

Even though the seeming constancy of $V_{33}$ for the three different charge states of the impurity quoted in Table II, the strong $V_{ii}$ variation as a function of the Cd charge state (displayed in Fig. 9), in particular near the completely ionized double acceptor state, justifies the existence of the randomly fluctuating EFGs mentioned in the proposed B-O model described in Sec. II A. On the other hand, the comparison between the predicted results quoted in Table II and Fig. 9 with the experimental EFG1 shows a perfect agreement only for the SnO$_2$:Cd$^{2-}$ system. This let us to definitively assigned HFI1 to completely ionized $^{111}$Cd probes localized at defect-free substitutional Sn sites.

The mean values of $V_{33}$ and $\eta$ of HFI2 [see Figs. 3(d) and 3(e)] are in good agreement with the predicted EFGs obtained when 1.5 up to 1.8 electrons are added to the SnO$_2$:Cd$^0$ SC. Again, this result let us to assign HFI2 to $^{111}$Cd probes, localized at defect-free substitutional Sn sites, with final stable charge states close to the completely ionized state of the acceptor impurity level.

Finally, taking into account the $V_{ii}$ behavior as a function of the charge state of the impurity and the proposed scenario within the B-O modelization, we showed that a slight time-dependent variation of the Cd charge state when it traps between 1.5 to 2 electrons is enough to give raise to the fluctuating hyperfine interactions observed in the TDPAC experiments.

## V. SUMMARY AND CONCLUSIONS

TDPAC experiments were performed on high purity rutile SnO$_2$ polycrystalline samples using $^{111}$Cd isotope as probe-atom as a function of measuring temperature in the range 10 K-1123 K in a reversible way. To introduce the radioactive $^{111}$Cd impurities in the samples, its parent $^{111}$In ($\rightarrow^{111}$Cd) was diffused under non oxidant N$_2$ atmosphere, aiming to substitute Sn atoms of the host lattice.

The TDPAC spectra showed a peculiar and increasing (reversible) damping as measuring temperature decreases that could be analyzed by a time-dependent perturbation factor in the framework of the Bäverstam and Othaz *"on-off"* model. Two hyperfine interactions, HFI1 and HFI2 were necessary to account for the spectra along all the temperature range of measurement. Both interactions are static at high temperatures and time-dependent at lower temperatures, but below a different temperature threshold for each interaction. The time-dependent perturbation factor results, in the B-O model, in a combination of a *dynamic* and a *static* term. While HFI1 is purely static for T ≥ 650K, HFI2 is static only for T ≥ 923 K. HFI1, characterized by EFG1, is very well defined (monochromatic), with its hyperfine parameters independent on temperature. HFI2, characterized by EFG2, presents a much larger EFG (static) distribution, being $V_{33}$ and $\eta$ oscillating parameters around a mean value larger than those of HFI1. With respect to the parameters that characterize the dynamic behavior, the relaxation constant $\lambda_r$, which measure the attenuation strength originated in the fluctuating EFGs, increases for both interactions as T decreases. This can be explained in terms of an increasing number of electronic holes as T decreases due to the decrease of the electron availability related with the thermal ionization of defects. The lifetime of the electronic holes ($1/\lambda_g$) increases as T decreases in the case of HFI1 and has a rather constant large value in the case of HFI2. In our scenario, the large value of $\tau_g$(HFI2) is due to the lack of enough electrons to fill the double acceptor level of the fraction of $^{111}$Cd atoms that feel HFI2. On the other hand, the behavior of $\tau_g$(HFI1) can be explained as follows: just above 650 K, $\tau_g$ is extremely short since there are enough number of electrons available to the rapidly ionized the small number of higly diluted $^{111}$Cd atoms that feels HFI1 at this temperature. As T decreases, $\tau_g$ increases due to the loss of thermal-excited electrons that could completely ionized the acceptor level.

In the B-O model applied to interpret our experimental results, EFG1 and EFG2 are related with different final stable electronic configurations of the $^{111}$Cd atom once the dynamic hyperfine interactions are *turn off*. The dynamic regime sensed by the $^{111}$Cd nucleus is originated in the random fast fluctuations between different electronic configurations of the $^{111}$Cd atom before arriving to the stable configurations that give raise to EFG1 and EFG2. Those configurations are correlated to different fluctuating charge states of the impurity (including those corresponding to the final stable configurations).

The detailed first-principles study of the electronic structure of Cd-doped $SnO_2$ showed that the Cd impurity introduces a double acceptor level in the top of the valence band and produces isotropic outward relaxations of the nearest oxygen neighbors that have to be taken into account self-consistently in order to correctly predict the hyperfine properties at the Cd site. The acceptor level has mainly contributions from Cd-$d$ and ONN-$p$ orbitals. In effect, when two electrons are added to the *neutral* Cd-doped $SnO_2$ SC, the acceptor level became completely filled, and hence this electronic charge contributes to enhance the electron density almost only at Cd and its ONN atoms. While the strength of $V_{33}$ is mainly due to the Cd-$p$ contribution (which is practically constant for the three $V_{ii}$ components) present along all the valence band, its variation and that of the asymmetry parameter $\eta$ as a function of the charge state of the impurity are due to the strong variation of the Cd-$d$ contribution to $V_{ii}$, which in turn is related to the filling of the acceptor level, obtained adding from 0 to 2 electrons to the SC. This strong variation of the EFG was explained in terms of the variation in the occupancy of the Cd-$d$ orbitals present in the acceptor level and the change in the anisotropy of the electronic density at the closest neighborhood of the Cd nucleus related with these orbitals.

The comparison between the *ab initio* predictions and the experimental EFG results enabled to assign HFI1 to $^{111}$Cd probes localized at defect-free cation sites of the rutile structure with a double acceptor level completely filled above T=650 K. Below this temperature, these probe-nuclei feel a dynamic hyperfine interaction whose final stable electronic configuration is that corresponding to the static EFG1 sensed above T=650 K. This EFG1 are in perfect agreement with the EFG that characterizes the single hyperfine interaction observed by Wolf *et al.*,[21] with the high-temperature interaction found by Bibiloni *et al.*,[22] and with the interaction assigned to $^{111}$Cd at Sn sites of the high-temperature crystalline phase in $SnO_2$ thin films.[24] On the other hand, HFI2 were assigned to $^{111}$Cd probes localized at defect-free Sn sites with the double acceptor level partially filled (between 1.5 and 1.8 added electrons) above T=923 K (static regime). Below this temperature, these probe-nuclei feel a dynamic hyperfine interaction whose final stable configurations (depending on T) is one of those corresponding to the impurity level charge states with 1.5 to 1.8 added electrons.

The dynamic behavior of HFI1 (at lower T) and the existence of an unexpected second HFI2 (in a host with a single cation site such as $SnO_2$) assigned to $^{111}$Cd probes that do not reach the completely ionization of the impurity acceptor state suggest that the parameters that characterize and give rise to the dynamic interactions depend on the insulating or semiconducting character of the host, hence on the electronic availability and mobility and, necessarily, on the variation of the EFG as a function of the charge state of the $^{111}$Cd impurity. The variation of the EFG in turn depends on the adopted configuration of the Cd orbitals upon electrons feeling (i.e., the resulting electronic density closest to the Cd nucleus) in each impurity-host system.

At this point we can conclude that the EC as well as the double acceptor character of the $^{111}$Cd impurity contribute to the existence of the dynamic hyperfine interactions since they promote the presence of electronic holes localized at the $^{111}$Cd atom with a lifetime long enough to be experimentally detected, depending on the conducting nature of the host. The same conclusion could be applied to $^{111}$In($\rightarrow$Cd)-doped In-, Y-, and Sc-sesquioxides[39,61,62] in which dynamic hyperfine interactions were also observed, being $^{111}$Cd a single acceptor in these host. Up to our knowledge, the ECAE seems to be a necessary condition to the observation of this type of this type of time-dependent interactions in oxides, taking into account two facts: first, that up to now no time-dependent interactions like those described here have been observed in PAC experiments using $^{181}$Hf($\rightarrow$$^{181}$Ta) in oxides, probe that does not decay through EC; second, when $^{111m}$Cd($\rightarrow$$^{111}$Cd) was used in PAC experiments in some of the same oxides mentioned above, dynamic hyperfine interactions were not observed or its effect was very weak.[43,63,64] At present, it is early to confirm that the electronic acceptor character of the Cd probe-atom is a necessary condition to observe dynamic interactions. For this, investigations in other binary oxides in which Cd would be an isovalent or a donor impurity in the host system should be perform. PAC experiments as well as *ab initio* calculations in this sense are currently in progress.

**ACKNOWLEDGEMENTS**

This work was partially supported by Consejo Nacional de Investigaciones Científicas y Técnicas (CONICET, under Grants No. PIP0002 and PIP0803), Conselho Nacional de Desenvolvimento Científico e Tecnológico (CNPq, grant no. 305046/2013-6), and Fundação de Amparo a Pesquisa do Estado de São Paulo (FAPESP, grant no.

2014/14001-1). This research made use of the computational facilities of the Physics of Impurities in Condensed Matter (PhI) group at IFLP and Departamento de Física (UNLP). Prof. Dr. A.F. Pasquevich is gratefully acknowledged for very fruitful discussions about the time-dependent perturbation factor and the ECAE phenomena. E.L.M. wishes to thank the personnel at IPEN for their kind hospitality during his research stay. M.R. and G.N.D. are members of CONICET, Argentina.

\* darriba@fisica.unlp.edu.ar

† renteria@fisica.unlp.edu.ar

|  |  | d Cd-O1 (Å) | d Cd-O2 (Å) |
|---|---|---|---|
| $SnO_2$ [a] |  | 2.047 | 2.057 |
| $SnO_2$:$Cd^0$ | LDA | 2.112 | 2.147 |
|  | WC-GGA | 2.110 | 2.149 |
|  | PBE-GGA | 2.110 | 2.151 |
| $SnO_2$:$Cd^{1-}$ | LDA | 2.132 | 2.161 |
|  | WC-GGA | 2.127 | 2.165 |
|  | PBE-GGA | 2.126 | 2.168 |
| $SnO_2$:$Cd^{2-}$ | LDA | 2.158 | 2.203 |
|  | WC-GGA | 2.153 | 2.204 |
|  | PBE-GGA | 2.152 | 2.207 |

[a]Reference 45.

**Table I:** Experimental Sn-O1 and Sn-O2 bond-lengths for pure $SnO_2$ (first line) and predicted equilibrium Cd-O1 and Cd-O2 distances for $SnO_2$:$Cd^0$ $SnO_2$:$Cd^{1-}$ and $SnO_2$:$Cd^{2-}$, using the LDA, WC-GGA, and PBE-GGA approximations.

|  |  | $V_{33}$ ($10^{21}$ V/m$^2$) | $\eta$ | $V_{33}$ direction |
|---|---|---|---|---|
| $SnO_2$:$Cd^0$ | LDA | +5.24 | 0.46 | [1-10] |
|  | WC-GGA | +5.15 | 0.40 | [1-10] |
|  | PBE-GGA | +5.11 | 0.36 | [1-10] |
| $SnO_2$:$Cd^{1-}$ | LDA | +6.17 | 0.59 | [1-10] |
|  | WC-GGA | +5.97 | 0.52 | [1-10] |
|  | PBE-GGA | +5.91 | 0.50 | [1-10] |
| $SnO_2$:$Cd^{2-}$ | LDA | +5.43 | 0.14 | [1-10] |
|  | WC-GGA | +5.67 | 0.20 | [1-10] |
|  | PBE-GGA | +5.75 | 0.19 | [1-10] |
| Exp. HFI1 |  | 5.68(1) | 0.13(1) | ----- |

**Table II:** $V_{33}$, $\eta$, and $V_{33}$ direction predicted for $SnO_2$:$Cd^0$, $SnO_2$:$Cd^{1-}$, and $SnO_2$:$Cd^{2-}$ using the LDA, WC-GGA, and PBE-GGA approximations. The experimental values of HFI1 are shown for comparison.

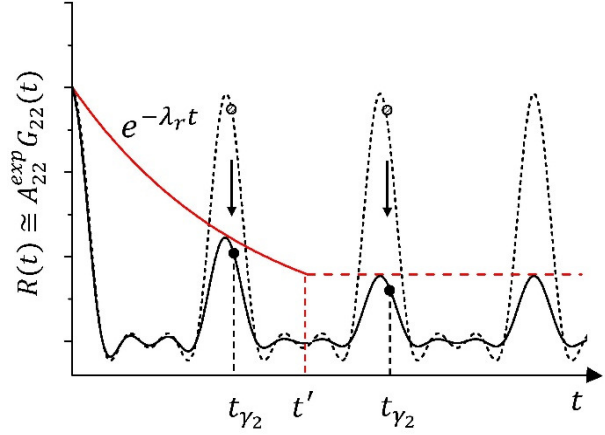

FIG. 1. (Color online) Graphical naive explanation of the construction of an $R(t)$ spectrum corresponding to TDPAC probes sensing a time-dependent *on-off* hyperfine interaction. Black dashed line, $R(t)$ spectrum for an axially symmetric *static* electric-quadrupole interaction (characterized by $EFG_1$). Red solid line represents the *pure dynamic* interaction of Eq. 4, which is switch-off after a time $t'$ (red dashed line). Black solid line, contribution to the total time-dependent R(t) spectrum (see Eq. 9) from all the probes sensing a *dynamic* hyperfine interaction (with a final stable $EFG_1$) as described by Eq. 7, reaching these probe-atoms the same stable electronic charge state configuration at a time $t'$, emitting $\gamma_2$ at different times $t_{\gamma 2}$.

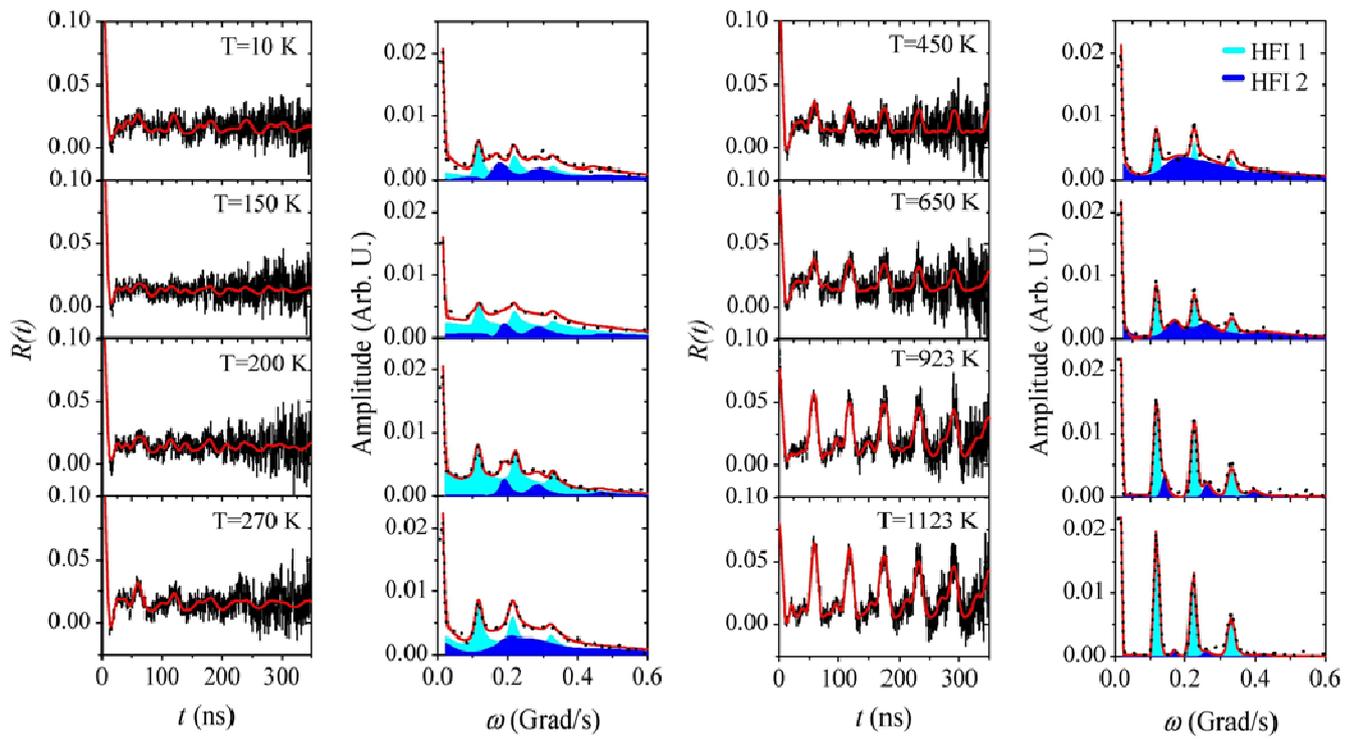

FIG. 2. (Color online) $R(t)$ spectra (left) and their corresponding Fourier transformed spectra (right) taken at the indicated selected measuring temperatures.

(Please insert as wide figure)

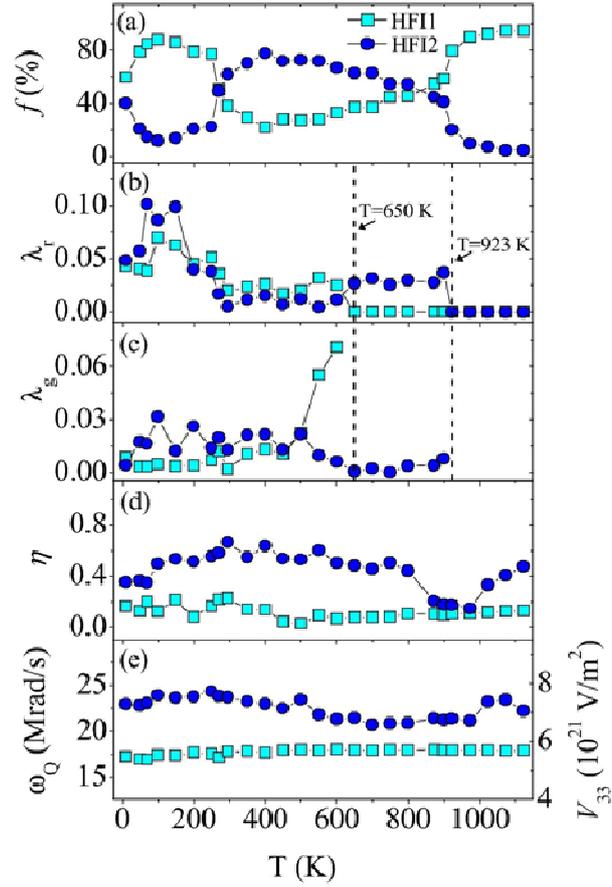

FIG. 3. (Color online) Evolution of (a) fractions and (b)-(e) hyperfine parameters $\lambda_r$, $\lambda_g$, $\eta$, and $\omega_Q$ (or $V_{33}$), of both hyperfine interactions HFI1 and HFI2, as a function of the measurement temperature T.

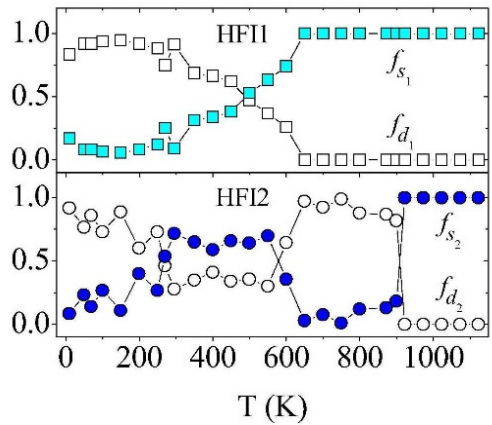

FIG. 4. (Color online) $f_s$ and $f_d$ for HFI1 (upper) and HFI2 (lower) hyperfine interactions, computed using Eq. 12 and the fitted $\lambda_r$ and $\lambda_g$ parameters shown in Fig 3(b) and 3(c).

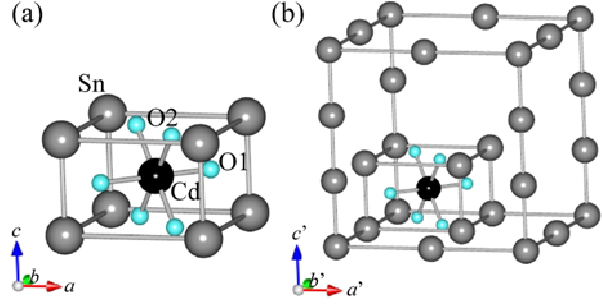

FIG. 5. (Color online) (a) Rutile $SnO_2$ unit cell. (b) Supercell with $a'=2a$, $b'=2b$ and $c'=3c$. The grey, black, and light-blue spheres correspond to Sn, Cd and O atoms, respectively.

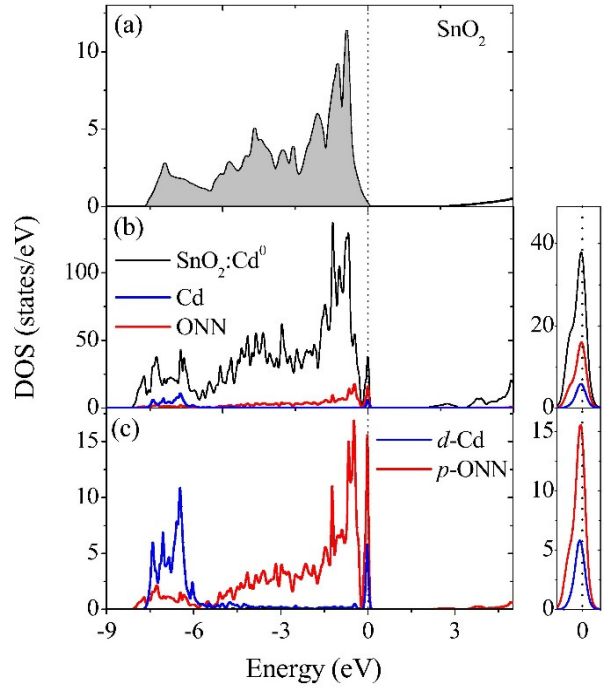

FIG. 6. (Color online) (a) Total density of states (DOS) for pure $SnO_2$. The shaded area shows the occupied states. (b) Total DOS (black) and atom resolved partial DOS (PDOS) projected at Cd (blue) and ONN (red) site for $SnO_2:Cd^0$ system. (c) PDOS of $d$ and $p$ contributions at Cd and ONN sites, respectively. The energy is referred to the last occupied stated (vertical dotted line). On the right of (b) and (c) panels we show a zoom of the impurity level region.

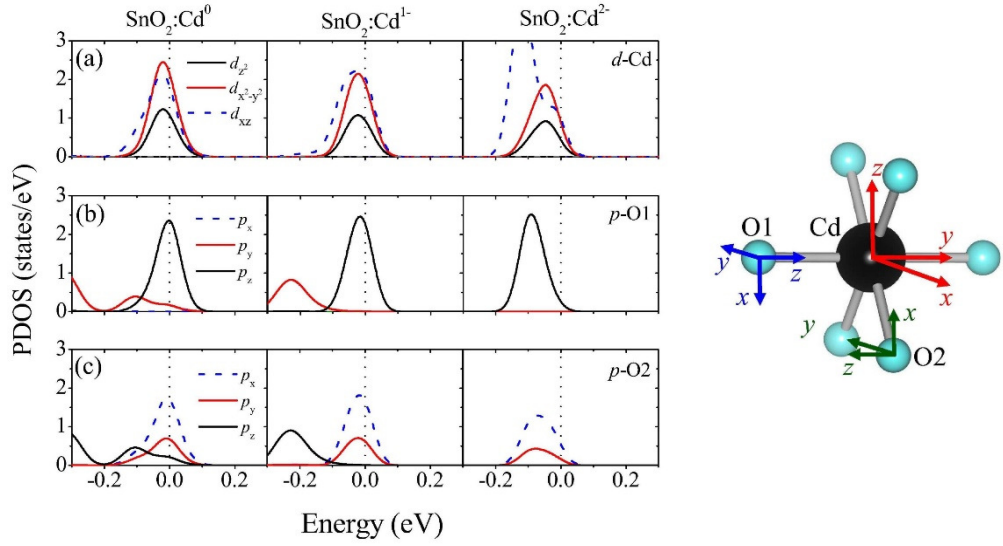

FIG. 7. (Color online) (a) PDOS of *d* contributions projected at Cd site, (b) and (c) PDOS of *p* contributions projected at O1 and O2 sites, respectively, for $SnO_2:Cd^0$, $SnO_2:Cd^{1-}$ and $SnO_2:Cd^{2-}$ systems. The energy is referred to the last occupied state (vertical dotted line). On the right, the local coordinate axes systems in which Cd-, O1-, and O2-projected PDOS were calculated.

(Please insert as wide figure)

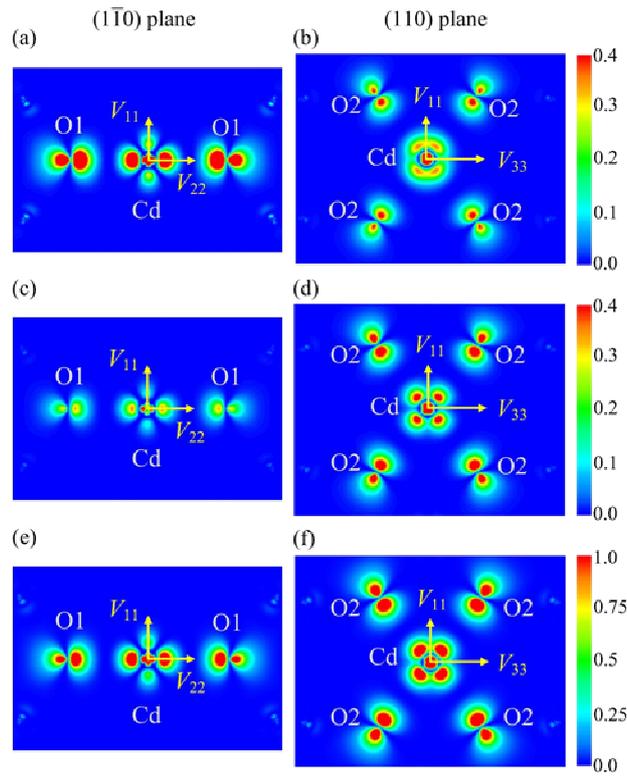

FIG. 8. (Color online) Electron density for (a) and (b) the first electron, (c) and (d) the second electron, and (e) and (f) the two electrons added to the SC, in the (1-10) (left column) and (110) (right column) planes.

("Centered figure", please enlarge Figure 8 conveniently)

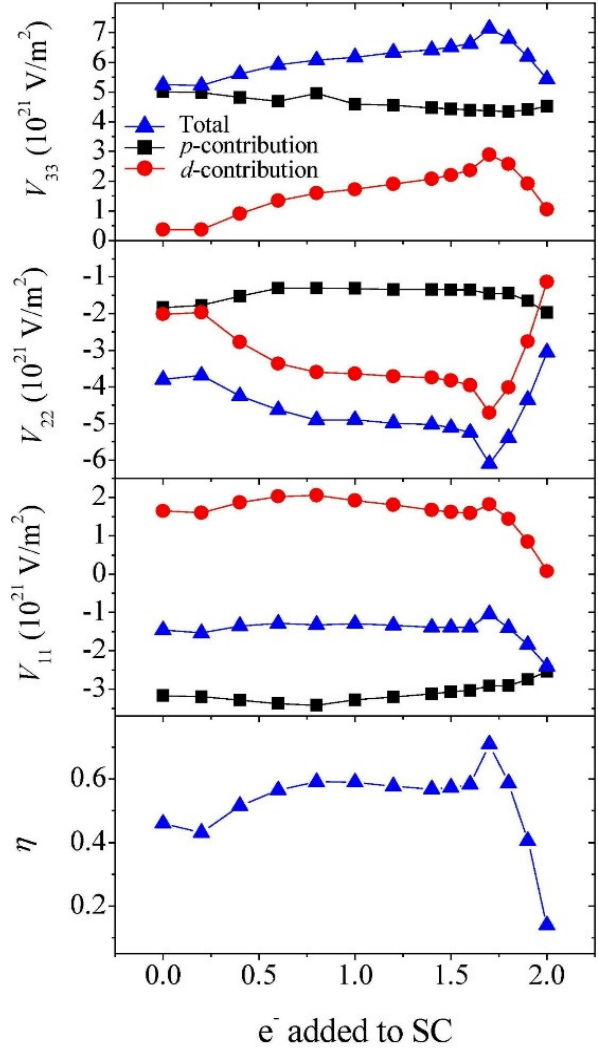

FIG. 9. (Color online) $V_{33}$, $V_{22}$, $V_{11}$, and $\eta$ parameter (blue triangles) as a function of the number of electrons added to the SC. For each $V_{ii}$, the $p$ and $d$ contributions (black squares and red circles, respectively) are shown.